\documentclass[aps,prd,preprint,superscriptaddress,nofootinbib,showpacs]{revtex4-1}

\usepackage{graphicx}% Include figure files
\usepackage{dcolumn}% Align table columns on decimal point
\usepackage{bm}% bold math
\usepackage{multirow}
\usepackage{color}%
\usepackage[english]{babel}
\usepackage{amsmath,amssymb}
\usepackage{amsfonts}
\usepackage{mathrsfs}
\usepackage{epsfig}
\usepackage{tabularx}
\usepackage{array}

\def\nuebar{{\rm \bar{\nu}_e}}
\def\nue{{\rm \nu_e}}
\def\numu{{\rm \nu_\mu}}
\def\numubar{{\rm \bar{\nu}_\mu}}
\def\nualphae{{\nu_\alpha e^-}}
\def\s2tw{{\rm sin ^2 \theta_{W}}}
\def\gbl{g_{\rm B-L}}
\def\aprime{{A^\prime}}

\begin{document}

% Use the \preprint command to place your local institutional report
% number in the upper righthand corner of the title page in preprint mode.
% Multiple \preprint commands are allowed.
% Use the 'preprintnumbers' class option to override journal defaults
% to display numbers if necessary
%\preprint{}

%Title of paper
%\title{Searching for Dark Photon with TEXONO Experiment}
\title{Constraints on Dark Photon from Neutrino-Electron Scattering Experiments}
% repeat the \author .. \affiliation  etc. as needed
% \email, \thanks, \homepage, \altaffiliation all apply to the current
% author. Explanatory text should go in the []'s, actual e-mail
% address or url should go in the {}'s for \email and \homepage.
% Please use the appropriate macro foreach each type of information

% \affiliation command applies to all authors since the last
% \affiliation command. The \affiliation command should follow the
% other information
% \affiliation can be followed by \email, \homepage, \thanks as well.
\newcommand{\as}{Institute of Physics, Academia Sinica, Taipei 11529, Taiwan.}
\newcommand{\metu}{Department of Physics, Middle East Technical University, Ankara 06531, Turkey.}
\newcommand{\dokuzeylul}{Department of Physics, Dokuz Eyl\"{u}l University, \.{I}zmir, Turkey.}
\newcommand{\bhu}{Department of Physics, Banaras Hindu University, Varanasi, 221005, India.}
\newcommand{\corr}{htwong@phys.sinica.edu.tw;
Tel:+886-2-2789-9682; FAX:+886-2-2788-9828.}

\author{ S.~Bilmi\c{s}}  \affiliation{ \metu }
\author{I.~Turan}  \affiliation{ \metu }
\author{T.M.~Aliev}  \affiliation{ \metu }
\author{ M.~Deniz }  \affiliation{ \as } \affiliation{ \dokuzeylul }
\author{ L.~Singh } \affiliation{ \as } \affiliation{\bhu}
\author{ H.T.~Wong } \affiliation{ \as }
%\author{ H.T.~Wong } \altaffiliation[Corresponding Author: ]{ \corr } \affiliation{ \as } 
%\collaboration{TEXONO Collaboration} \noaffiliation
%\email[]{Your e-mail address}
%\homepage[]{Your web page}
%\thanks{}
%\altaffiliation{}
%\affiliation{METU}

%Collaboration name if desired (requires use of superscriptaddress
%option in \documentclass). \noaffiliation is required (may also be
%used with the \author command).
%\collaboration can be followed by \email, \homepage, \thanks as well.
%\collaboration{}
%\noaffiliation

\date{\today}

\begin{abstract}
% insert abstract here
A possible manifestation of an additional light gauge boson $A^\prime$, named as Dark Photon, associated with a group $U(1)_{\rm B-L}$ is studied in neutrino electron scattering experiments.  The exclusion plot on the coupling constant $g_{\rm B-L}$ and the dark photon mass $M_{A^\prime}$ is obtained. It is shown that contributions of interference term between the dark photon and the Standard Model are important. The interference effects are studied and compared with for data sets from TEXONO, GEMMA, BOREXINO, LSND as well as CHARM II  experiments. Our results provide more stringent bounds to some regions of parameter space.
\end{abstract}

% insert suggested PACS numbers in braces on next line
\pacs{13.15.+g,12.60.+i,14.70.Pw}
% insert suggested keywords - APS authors don't need to do this
\keywords{neutrino, dark photon, hidden sector, texono}

%\maketitle must follow title, authors, abstract, \pacs, and \keywords
\maketitle

% body of paper here - Use proper section commands
% References should be done using the \cite, \ref, and \label commands
%\section{Introduction}
% Put \label in argument of \section for cross-referencing
%\section{\label{}}
 
\tableofcontents
 
\section{Introduction}
\label{vlvb}

The recent discovery of the Standard Model (SM) long-sought Higgs at the Large Hadron Collider is the last missing piece of the SM which is strengthened its success even further. Of course this does not change the fact that there are the issues of neutrino mass, the presence of dark matter etc. and thus the SM is an effective theory whose range of validity has to be tested in either direction from the weak scale. While the scale of new physics sets up one boundary at the higher end, the mass scale of the neutrinos could be considered one of the fundamental scales in physics at the lower tail around which the SM's validity should also be questioned. For instance neutrino nucleus coherent scattering has not been observed yet~\cite{Wong:2005vg}, which will test the SM at very low energies.

In the quest for new physics, the limitations of SM can be tested through high-energy frontiers as well as through intensity frontier with high-precision experiments which is considered to be complementary to the direct searches at high energies. There are numerous experimental results such as, the anomalous magnetic moment of the muon \cite{muon,Bennett:2006fi}, smallness of electric dipole moment of neutron \cite{muon,Baker:2006ts}, electric charge radius puzzle of proton~\cite{Pohl:2010zza}, the positron excess in cosmic rays without anti-proton abundance (first seen by ATIC experiment \cite{Chang:2008aa} and later confirmed by PAMELA \cite{pamela} and FERMI \cite{fermi} satellite  experiments) as well as INTEGRAL satellite experiment observation of a very bright 511keV line \cite{Jean:2003ci} together with other puzzling results coming out of DAMA/LIBRA \cite{dama} and EGRET \cite{egret} collaborations and also recent AMS-02 experiment announcement about the positron excess even with a sharper rise up to 300 GeV energies \cite{ams}, none of which can be explained within the SM. Hence, new physics scenarios beyond the SM are needed. Even though finding a way out to one or two of these is a step, the real ambitious challenge is to find a framework where all or at least most of all of these puzzling inconsistencies find themselves a remedy without violating any of the existing data.

As a remedy to some of these issues, we will consider a hidden sector scenario where the existence of a dark photon may alter significantly the neutrino-electron scattering data or at least its gauge coupling and mass could be constrained with the use of the data. There are various neutrino-electron scattering experiments which are mainly TEXONO~\cite{texononue,texonomunu,texonoNPCGe}, BOREXINO~\cite{Bellini:2011rx}, GEMMA~\cite{Beda:2009kx} as well as LSND~\cite{lsnd} and CHARM II~\cite{charm2}. A light dark photon could be searched using these data.
 
 The paper is organized as follow. In section \ref{model}, the idea of hidden sector and some details of the considered model will be described. In section \ref{nes}, the details of neutrino electron scattering in Standard Model as well as the $U(1)_{\rm B-L}$ dark photon scenario will be given. Pure dark photon as well as interference contributions to the differential cross sections of various neutrino electron scattering processes are presented.
 In section \ref{analysis}, our results are compared with the existing results in literature. Especially, the interference effects are discussed in detail and its importance for some cases is stressed. Section \ref{conc} contains our conclusions.

\section{Hidden Sector as a beyond the Standard Model Scenario}
\label{model}

The idea of existence of a so-called hidden sector interacting with the SM through various portals (more on portals is below) is one of such extensions of the SM aiming to explain some of the above issues. With a single particle from the hidden sector being singlet under SM gauge group, there is no way to couple with the visible part other than its gravitational effects   
which will be suppressed by the Planck scale, putting them out of reach of any current experimental search (it should then be called truly hidden sector). So for testable scenarios, more than one hidden sector fields should play a role in portal.

One may consider couplings of the form ${\cal L} = \sum_{l,m}{\cal O}_{\rm HS}^{(l)}{\cal O}_{\rm SM}^{(m)}$ where ${\cal O}_{\rm HS}\, ({\cal O}_{\rm SM})$ are some hidden (SM) sector operator  and if the sum of the dimensions of the operators is $l+m=4$, there will be no suppression due to high cutoff scale. Such SM operators at the lowest order are known as portals like vector portal, Higgs portal, neutrino portal, axion portal, etc.

Among many possible portals mentioned above, the so-called vector portal assumes a hidden sector vector boson coupled to the SM gauge boson(s) through a kinetic mixing which could be generated through one-loop by exchange of a heavy messengers having non-zero charges under both SM and hidden sector gauge groups. There are alternatives one can consider for the gauge group from the hidden sector but the simplest choice would be an abelian symmetry as extra $U(1)$, dubbed as $U(1)^\prime$,  which is well motivated from both the top-down (grand unification, string theory etc) and bottom-up (dark matter and other issues mentioned above) approaches in extending the SM to tackle with the puzzles at hand. 

With a $U(1)^\prime$ hidden sector gauge symmetry, it mixes with the corresponding SM $U(1)_Y$ in the same representation through a renormalizable operator by a kinetic-term mixing mechanism (this is a way to avoid otherwise strong theoretical and experimental constraints due to this new interaction). The hidden sector gauge field of $U(1)^\prime$ is called hidden or dark photon. The mixing parameter $\epsilon$ is constrained by the scale of the messenger fields. Further suppression occurs when the SM gauge group is embedded into a bigger grand unified picture in the top-bottom approach where the leading contributions would be two-loop. In the bottom-up approach breaking the $U(1)^\prime$ symmetry at very light scales is not very unusual since it seems that neutrino mass differences indicate existence of another fundamental scale in that regime. A non-zero but tiny mass needs to be considered to the new $U(1)^\prime$ gauge field since zero mass case is inconsistent with the current observations if the dark photon is further assumed to be a dark matter candidate.
 
Even though the idea of very light vector bosons from the hidden sector in the form of a dark photon, is not new \cite{Okun:1982xi, Holdom:1985ag}, their effects on various SM processes at low energies in intensity frontiers has recently received great attention, which might be partly due to lack of any new physics signal at Large Hadron collider.

The allowed interactions of dark photon with the SM particles depend on the theoretical framework. There are two main approaches the way to couple dark photon sector with the SM. One common practice is to make the dark photon mix with the photon through a kinetic mixing so that, like the SM photon, its coupling only with the charged fermions would be induced. The mass of dark photon and a kinetic mixing parameter are the only additional ingredients of the model. Note that even though the new gauge coupling constant is involved in the definition of the kinetic mixing parameter $\epsilon$ through one-loop diagram, it does not affect directly the dark photon coupling to the SM particles.

Another way to connect dark photon sector with the SM is through a $U(1)$ gauging, like $U(1)_{B-L}$, where the dark photon as the gauge field of the group interacts with any SM particle with non-zero $\rm B-L$ number at tree level. Here the new gauge coupling constant and the dark photon mass are the free parameters by ignoring the kinetic mixing. Even though considering these one-at-a-time basis is mostly adopted in order to have a better predictability power, there is no prior reason not to allow both at the same time. Our aim is to bound the coupling constant $g_{B-L}$ directly rather than translating the bound on $\epsilon$.

Let us consider the Lagrangian including both the kinetic mixing with the hypercharge $U(1)_Y$ and the $B-L$ coupling. We have
\begin{widetext}
\begin{eqnarray}
\label{lnotrot}
{\cal L} = -\frac14 B^{\prime 2}_{\mu\nu} - \frac14 F^{\prime\prime 2}_{\mu\nu}+\frac12\epsilon^\prime B^{\prime}_{\mu\nu}F^{\prime\prime\mu\nu}+\frac12 M_{A^{\prime\prime}}^2 A^{\prime\prime 2}_\mu + g_Y j_B^\mu B^{\prime}_\mu + g_{B-L} j_{B-L}^\mu A^{\prime\prime}_\mu + ...
\end{eqnarray}
where $B^\prime_\mu$ and $A^{\prime\prime}_\mu$ are the gauge fields of $U(1)_Y$ and $U(1)_{B-L}$ groups, respectively and the currents are defined as
	\begin{eqnarray}
j_{B}^\mu &=& \frac{e}{g_Y}\left(\cos\theta_W j_{\rm em}^\mu -\sin\theta_W j_Z^\mu \right)\nonumber\\
j_{B-L}^\mu  &=& (B-L)\bar{f}\gamma^\mu f = -\bar{\ell}\gamma^\mu \ell - \bar{\nu}_\ell \gamma^\mu \nu_\ell +\frac13 \bar{q}\gamma^\mu q.\nonumber
	 \end{eqnarray} 
\end{widetext}
The kinetic mixing can be eliminated by rotating the fields from $(B^\prime_\mu,A^{\prime\prime}_\mu)$  to $(B_\mu,A^\prime_\mu)$ as given first order in $\epsilon^\prime$, $B^{\prime\prime}_\mu \simeq B_\mu + \epsilon^\prime A^\prime_\mu\,,\;\; A^{\prime\prime}_\mu \simeq A^\prime_\mu$, and we get
	\begin{eqnarray}
	\label{lrot}
	{\cal L} = -\frac14 B^2_{\mu\nu} - \frac14 F^{\prime 2}_{\mu\nu}+\frac12 M_{A^\prime}^2 A^{\prime 2}_\mu+ g_Y j_B^\mu B_\mu + g_{B-L} j_{B-L}^\mu A^{\prime}_\mu + e\epsilon j_{\rm em}^\mu A^\prime_\mu + ...
	\end{eqnarray}
where $M_{A^{\prime\prime}} \simeq M_{A^{\prime}}$ and $\epsilon\equiv \epsilon^\prime \cos\theta_W$. The original kinetic mixing term in Eqn.~(\ref{lnotrot}) turns into the last term in Eqn.~(\ref{lrot}) which represents interaction of dark photon with charged matter field with coupling $e\epsilon$. Since without the  $\rm B-L$ gauging the dark photon does not couple with the neutrinos at tree level, we prefer to consider $\rm B-L$ and set the kinetic mixing zero.  We will focus on the searching of dark photons with neutrino experiments which has the advantage of being purely leptonic process.

\section{Neutrino-Electron Scattering}
\label{nes}

\subsection{Standard Model Expressions}
\begin{figure}[htb]
%	\vskip 0.3cm
%	\centering   
		\includegraphics[scale=0.8]{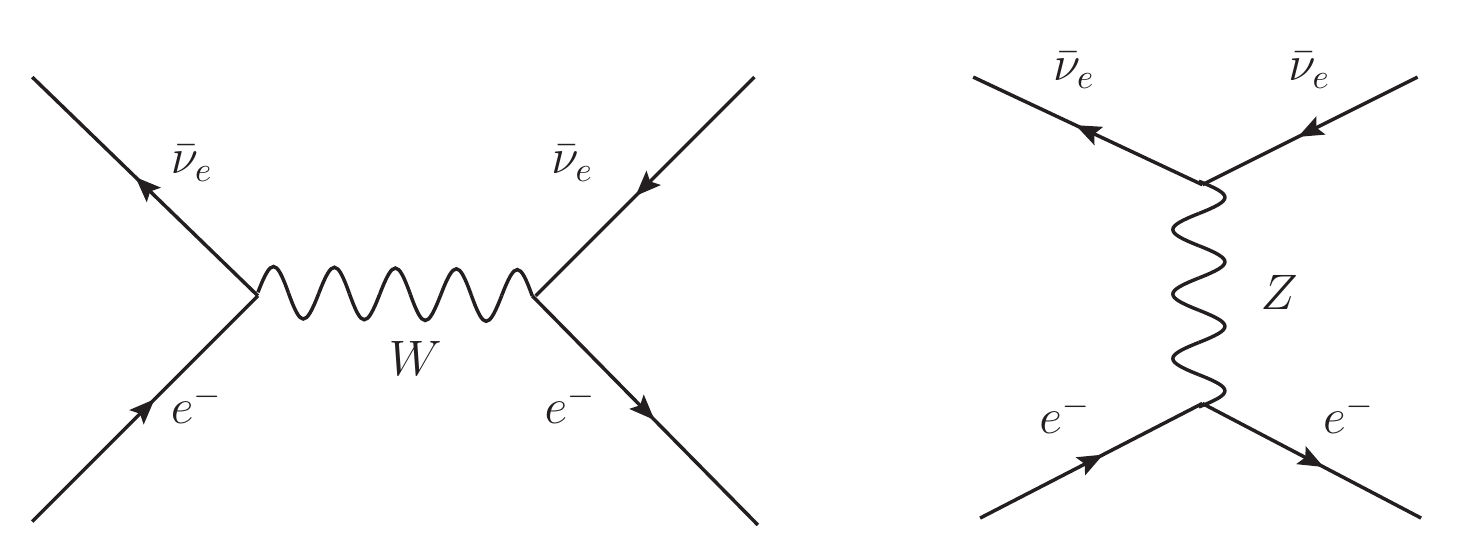} 	
	\caption{Electron neutrino electron scattering interaction takes place via both charged and neutral currents. For neutrinos other than the electron type, only the neutral current is involved.}
		\label{smfeyn}
\end{figure}

Neutrino interactions are purely leptonic processes with robust SM predictions. Hence searching physics beyond the SM in neutrino electron scattering turns out to be good alternative to collider searches. In the SM, the $\nue - e$ scattering takes place via both charged and neutral currents. However the $\nualphae$ scattering in which $\alpha$ corresponds to $\mu$ or $\tau$ occurs only due to neutral current. (See Fig.~\ref{smfeyn} for the relevant diagrams.) 

The differential cross-section in lab frame of the electron in Standard Model can be written as

\begin{equation}
\Big[\frac{d\sigma}{dT}(\nu  e^-\to \nu e^-)\Big]_{\rm SM}  = \frac{2 G_F^2 m_e}{\pi E_\nu^2}\Big(a^2 E_\nu^2+b^2(E_\nu-T)^2-a b m_e T\Big)\,,
\label{smcs}
\end{equation}
where $G_F$ is the Fermi coupling constant, $T$ is the recoil energy of the electron, $E_\nu$ is the energy of the incoming neutrino and $m_e$ is the mass of the electron. 
The differential cross sections differ depending on the neutrino flavour, i.e depending on parameters $a$ and $b$. The values of a and b are given in Table~\ref{SM_cross_sections}.
The maximum recoil energy of the electron depends on the mass of the electron as well as incoming neutrino energy as
$$T_{max} = \frac{2 E_\nu^2}{m_e + 2E_\nu}\,,$$
which also means that minimum neutrino energy required to give the electron a recoil energy $T$ is
\begin{eqnarray}
\label{Enumin} E_{\nu_{min}} = \frac{1}{2} T+\sqrt{T^2 + 2 T m_e}\,.
\end{eqnarray}

Any deviation of the recoil energy spectra of electron from what the SM predicts could be taken as a smoking gun for new physics. Our earlier works include studies of non-standard interaction parameters as well as unparticle and non-commutative physics~\cite{nsi,Bilmis:2012sq}. The dark photon contributions as well as its interference effects with the Standard Model are explored in the next section.

%*****************************************************
\begin{center}
	\begin{table}[t] 
		\vskip 0.3cm
		\caption{The parameters $a$ and $b$ in the SM cross section expression in Eqn.~(\ref{smcs}).}
		\begin{tabularx}{0.95\textwidth}{l@{\hskip 1.8in}l@{\hskip 2.0in}l}
			\hline\hline
			Process  &  $a$  & $b$  \\   
			\hline
			$ \nue e^- \rightarrow \nue e^- $ & $  \s2tw + \frac{1}{2} $ & $ \s2tw $  \\
			%\hline
			$ \nuebar e^- \rightarrow \nuebar e^- $& $ \s2tw $ & $ \s2tw + \frac{1}{2} $   \\
			%\hline
			$\nu_\alpha e^- \rightarrow \nu_\alpha e^- $ &  $  \s2tw -\frac{1}{2} $ & $\s2tw $  \\
			%\hline
			$ \bar{\nu}_\alpha e^- \rightarrow \bar{\nu}_\alpha e^-$ & $ \s2tw $ & $  \s2tw -\frac{1}{2} $  \\
			\hline\hline
		\end{tabularx}
		\label{SM_cross_sections}
	\end{table}
\end{center}

\subsection{Very Light Vector Boson Contributions}
\begin{figure}[t]
	\begin{tabular}{@{}c@{\hskip 1.0in}c@{\hskip 1.0in}c@{}}    
		\includegraphics[width=3cm]{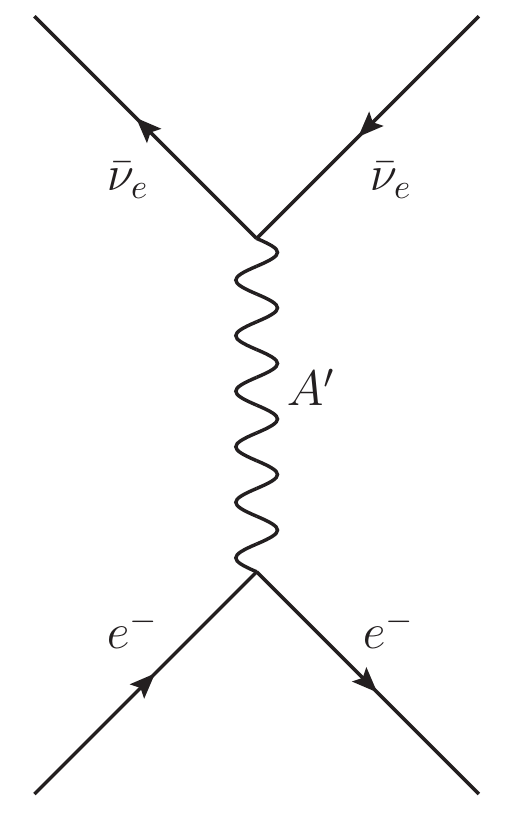} &
		\includegraphics[width=3cm]{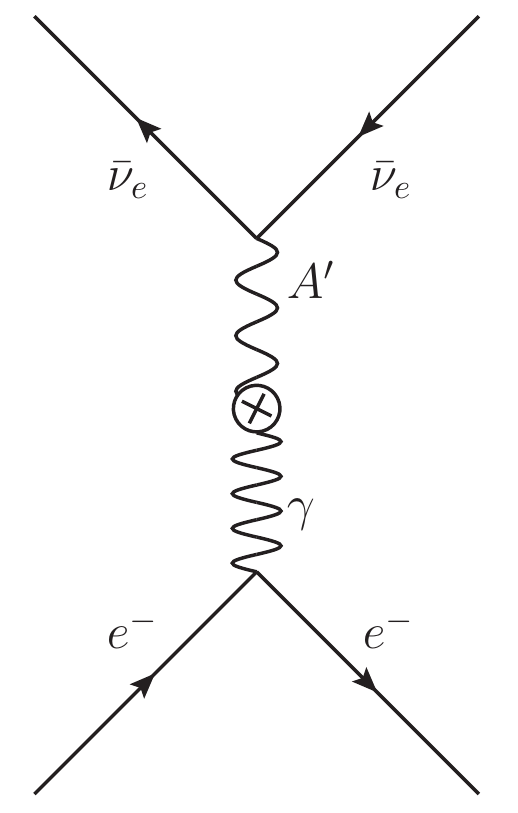} &
		\includegraphics[width=3cm]{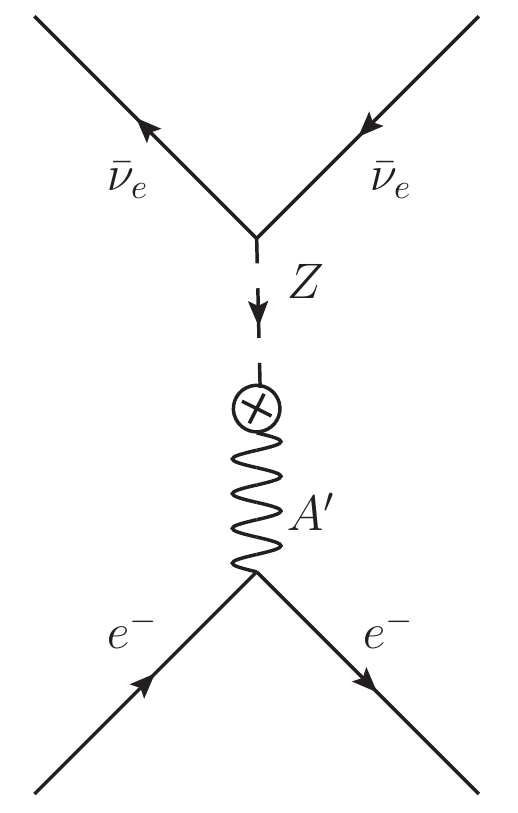} \\
		(a)         &  (b)  &  (c)
	\end{tabular}
	\caption{\label{Aprime} Interactions of neutrinos with electron via $t$ channel dark photon $A^\prime$ exchange in panel (a). The panels (b) and (c) are for the kinetic mixing between photon-dark photon and Z boson-dark photon, respectively. }
\end{figure}

Now let us calculate the contributions of the new light vector boson to the neutrino electron scattering processes. But first few comments are in order. The general form of the renormalizable Lagrangian given in Eqn. (\ref{Aprime}) where the dark and conventional photons can be mixed via kinetic term as mentioned earlier.  Analyses of the current experimental results lead to the maximum value of the mixing parameter $\epsilon$ of the order $10^{-2}$~\cite{Essig:2013lka}. This mixing has been extensively  studied in the literature (see~\cite{Essig:2013lka,Essig:2010gu,Essig:2009nc} and references therein). $B-L$ gauged $U(1)^\prime$ hidden sector scenario will also have a gauge coupling $g_{\rm B-L}$ as a free parameter in addition to its mass $m_{A^\prime}$ and $\epsilon$.  

As mentioned in the previous section, even though one can consider all three parameters ($M_{A^\prime},\, \epsilon,\, g_{\rm B-L}$) to do a fit to the data, in the present work, we will focus on a model with only two free parameters $M_{A^\prime}$ and $g_{\rm B-L}$ and ignore the effect of kinetic mixing.  Such analysis has not been done for experiments like TEXONO, LSND, or CHARM II. For the BOREXINO and GEMMA, there is a study~\cite{Harnik:2012ni} without considering the interference effects. There are other studies using a broken \cite{Williams:2011qb} and unbroken \cite{Heeck:2014zfa} $U(1)_{\rm B-L}$ scenarios to discuss neutrino-electron scattering.

Let us mention what is new in this study. First of all, the importance of interference effects which is overlooked in the literature is discussed. Our results show that interference effects are not always negligible and can enhance the results as large as one order for some cases. Second, we obtained bounds on $g_{\rm B-L}$ without relating it through the bound on the kinetic mixing parameter $\epsilon$. For this purpose $\epsilon$ parameter is not considered at all.  Third, the analyses for the TEXONO, LSND and CHARM II data have been done for the first time, and we repeat analyses for GEMMA and BOREXINO and found out that, unlike GEMMA case, the bound on $\gbl$ gets better for the BOREXINO when the interference effects are included.

After this preliminary remarks, let calculate contributions of light dark photon to the neutrino electron scattering processes. (See Fig.~\ref{Aprime}) Note that the diagrams Fig.~\ref{Aprime}b and \ref{Aprime}c would exist only when there is a kinetic mixing between the dark photon and the SM neutral gauge bosons. Thus, such contributions are ignored.

The pure contribution of this new diagram to the neutrino electron scattering is calculated and the differential cross section is obtained as 
\begin{equation}
\label{pure_DP_cs}
%\displaystyle
\Big[\frac{d\sigma}{dT} (\nu e^- \to \nu e^-)\Big]_{\rm DP} = \frac{g_{B-L}^4 m_e}{4\pi E_\nu^2 (M_{A^\prime}^2 + 2m_e T)^2}\Big(2 E_\nu^2 + T^2-2 T E_\nu - m_e T\Big) \,,
\end{equation}
where the cross section is neutrino flavor blind\footnote{The analytical expressions for the differential cross section in the SM, pure DP as well as including the interference cases are manually calculated and double checked with the package program $\tt CalcHEP$ \cite{Belyaev:2012qa}.}. For concreteness it is assumed that $\aprime$ has pure vector couplings of the form $\bar{f} \gamma^\mu f A^\prime_\mu$. 
For deriving the cross section formula, neutrinos are assumed to be massless. One of the key point in this study is to calculate and discuss the effect of interference. Our analysis has shown that the interference of this gauged B-L model with the SM can not be neglected for at least partly and should have been taken into account as opposed to the Ref.~\cite{Harnik:2012ni}. We discuss the criteria when the interference effects become sizable.

By using the diagrams given in Fig.~\ref{smfeyn} and \ref{Aprime}a, the interference differential cross section for each neutrino channel are obtained as

%\begin{widetext}
	\begin{eqnarray}
	\hspace*{-2cm}	\frac{d\sigma_{\rm INT}(\nu_e e^-)}{dT}  &=& \frac{\gbl^2 G_F m_e}{2 \sqrt{2} E_\nu^2 \pi (M_{A^\prime}^2 + 2 m T)}  \Big(~ 2 E_\nu^2 - m_e T   + \beta \Big)\,,\label{sigmanue}\\ 
	%	\frac{d\sigma_{\rm INT}(\bar{\nu}_\alpha \rightarrow e^-)}{dT}  &=& \frac{\gbl^2 G_F m_e}{2 \sqrt{2} E_\nu^2 \pi (M_{A^\prime}^2 + 2 m T)} \big[~ -2 E_\nu^2 - 2 T^2  + T(4 E_\nu +m_e) +  \s2tw (8 E_\nu^2 - 8 E_\nu T -4 m_e T + 4 T^2)~\big]\,, \\
	\frac{d\sigma_{\rm INT}(\nuebar e^-)}{dT} &=& \frac{\gbl^2 G_F m_e}{2 \sqrt{2} E_\nu^2 \pi (M_{A^\prime}^2 + 2 m T)}  \Big(~ 2 E_\nu^2 + 2 T^2  - T (4E_\nu + m_e) +  \beta ~\Big)\,, \label{sigmanubare}\\
	\frac{d\sigma_{\rm INT}(\nu_\alpha e^-)}{dT} &=& \frac{\gbl^2 G_F m_e}{2 \sqrt{2} E_\nu^2 \pi (M_{A^\prime}^2 + 2 m T)}   \Big( -2 E_\nu^2 + m_e T   + \beta~\Big)\,, \\
	\frac{d\sigma_{\rm INT}(\bar{\nu}_\alpha e^-)}{dT}  &=& \frac{\gbl^2 G_F m_e}{2 \sqrt{2} E_\nu^2 \pi (M_{A^\prime}^2 + 2 m T)} \Big( -2 E_\nu^2 - 2 T^2  + T(4 E_\nu +m_e) + \beta~\Big)\,, 
	\label{sigmaint}	
	\end{eqnarray}	
%\end{widetext}
where the parameter $\beta$ is defined as $$\beta = \s2tw (8 E_\nu^2 - 8 E_\nu T -4 m_e T +4T^2).$$

The index $\alpha$ in $\nu_\alpha$ is either $\mu$ or $\tau$ and they are different from the electron neutrino case since only $Z$ boson exchange diagram contributes in the former case while both $Z$ and $W$ bosons exchange diagrams contribute in the latter. A detailed analysis of the interference effects will be given in the next section. 

\section{Experimental Constraints}
\label{analysis}

\subsection{Neutrino-Electron Scattering Experiments}

Neutrino scattering experiments are good place for searching light dark photon. As seen from Eqn.~(\ref{pure_DP_cs}), for the low mass region of $M_{A^\prime}$ and for lower recoil energies of the electron, the differential cross section increases, which motivates to search new physics under such circumstances. Thus, experiments looking for dark matter particles or neutrino magnetic moment, which requires low recoil energies, are good places to search these effects. Among them, for example, the TEXONO Collaboration in Taiwan has various set of experiments each of which is designed for different physics purposes with  different recoil energy coverage. 

These recoil ranges as well as the average incident neutrino energies and the corresponding measured $\sin^2\theta_W$ values are summarized in Table~\ref{tab::experiments}. The Table also includes the information for similar experiments like LSND, BOREXINO, GEMMA, and CHARM II. Since, for the larger mass values of $M_{A^\prime}$, experiments with higher recoil energies of the electron with energetic neutrino source will also be effected. This motivates to search for the dark photon effects in the  neutrino sector by using the LSND and CHARM II experiments which measured the $\s2tw$ with the process $\nue$ and $\numu (\numubar)$ scattering respectively. 
\begin{table*} [t]
	\caption{The key parameters of
		the TEXONO, LSND, CHARM II, BOREXINO and GEMMA measurements 
		on the $\nu - e$ scattering.}
	\label{tab::experiments}
	\begin{ruledtabular} 
		\begin{tabular}{lccccc}
			Experiment  &  
			Type of neutrino & $\left< E_{\nu} \right>$ & $T$ & Measured $\s2tw$ 
			\\ \hline 
			TEXONO-NPCGe~\cite{texonoNPCGe} & $\nuebar$ &  1$-$2~MeV & 0.35$-$12 keV & $-$ 
			\\
			TEXONO-HPGe~\cite{texonomunu} & $\nuebar$ & 1$-$2~MeV & 12$-$60~keV & $-$ 
			\\
			TEXONO-CsI(Tl)~\cite{texononue} & 
			$\nuebar$ & 1$-$2~MeV & 3$-$8~MeV &0.251 $\pm$ 0.039 
			\\
			LSND~\cite{lsnd} &
			$\nue$ & 36~MeV &18$-$50~MeV & 0.248 $\pm$ 0.051
			\\
			BOREXINO~\cite{Bellini:2011rx} &
			$\nue$ & 862 keV & 270$-$665~keV & $-$ 
			\\
			GEMMA~\cite{Beda:2009kx} &
			$\nuebar$ & 1$-$2~MeV & 3$-$25~keV & $-$ 
			\\
			CHARM II~\cite{charm2} &  
			$\nu_{\mu}$ &23.7~GeV &3-24~GeV &
			\multirow{2}*{{\Huge \}} \vspace*{0.1cm} 0.2324 $\pm$ 0.0083 } &
			\\
			& $\bar{\nu}_{\mu}$ &19.1~GeV &3-24~GeV  & 
			\\
		\end{tabular}
	\end{ruledtabular}
	%	\label{tab::results}
\end{table*}

A brief summary of the experiments listed in Table~\ref{tab::experiments} would be useful here. The first in the list is the TEXONO experiment. TEXONO Collaboration has a research program on low energy neutrinos conducted at Kuo-Sheng Neutrino Laboratory which is located at a distance of 28 m from one of the cores of Kuo-Sheng Nuclear power station in Taiwan. Note that TEXONO is a reactor neutrino experiment with the advantage of high neutrino flux, hence mean energy of neutrinos is  $\left<E_\nu\right>=1-2$ MeV with a flux $\rm 6.4 \times 10^{12}$ cm$^{-2}$s$^{-1}$.

Three different data sets of TEXONO, each of  which are used for different purposes, have been analyzed. Let us summarize them below.
\begin{enumerate}
	\item \textbf{CsI:} A total mass of 187 kg CsI(Tl) crystal array is used to measure $\nuebar - e^-$ cross section with 29882/7369 kg-day of reactor ON/OFF data. Analysis range for recoil energy of electron is 3-8 MeV and the Weinberg angle is measured with the data (see Table~\ref{tab::experiments}).
	
	\item \textbf{HPGe:} Limits are set to neutrino magnetic moment with a target mass of 1.06 kg HpGe detector. 570.7/127.8 kg day of reactor ON/OFF exposure is taken and 10 keV analysis threshold with $\rm \sim 1kg^{-1}~keV^{-1}~day^{-1}$ background is achieved.
	
	\item \textbf{NPCGe:} N-type point-contact germanium detector of 500 g fiducial mass with 124.2 days reactor ON and 70.3 days reactor OFF data. $0.3-$keV threshold, used in previous search of neutrino milli-charge~\cite{texonoNPCGe}.
\end{enumerate}  

Unlike TEXONO, LSND (Liquid Scintillator Neutrino Detector) is a $\nue-e^-$ scattering experiment in which accelerator neutrinos are used as a source. Electron neutrino  beams are produced by decaying of $\mu^+$ at rest at Los Alamos Neutron Science Center with a mean energy $\left< E_\nu\right> \simeq 36$ MeV and total flux $\rm 11.76 \times 10^{13}\, cm^{-2}$. Analysis range for the recoil energy is $T \simeq 18-50$ MeV. The measured Weinberg angle is depicted in Table~\ref{tab::experiments} by using a sample of $191 \pm 22$ events.

CHARM II Collaboration measured electroweak parameters (see Table~\ref{tab::experiments}) using $\numu$ and $\numubar$ electron scattering based on $2677\pm 82$ and $2752 \pm 88$ events respectively. Neutrino beams are acquired via decay of pion at CERN. Mean energy of $\left<E_\numu\right> = 23.7$ GeV and $\left<E_\numubar\right> = 19.1$ GeV. The energy range of the analysis is $3-24$ GeV.

BOREXINO Collaboration measured the spectrum of $\rm ^7Be$ solar neutrino (with 862 keV energy) via elastic scattering of neutrinos using liquid scintillator. Analysis range for the  recoil energy of electron is $270-665$ keV.

GEMMA Collaboration measured the neutrino magnetic moment with data taken for three years using 1.5 kg HPGe detector with an energy threshold 3 keV. The experiment is located et the Kalinin Nuclear Power Plant in Russia. Therefore $\nuebar$ is used as a source with the energy $\left<E_\nu\right> \sim 1-2$ MeV and $\nuebar$ flux is $2.7 \times 10^{13}$ cm$^{-2}$s$^{-1}$. For the analysis, 13000 ON-hours and 3000 OFF-hours of data are used.

Having shortly mentioned the experiments, let us summarize the procedure used in the analysis. The contribution of the dark photon to the electron recoil spectra is calculated as 
\begin{equation}
\frac{dR_{\rm DP}}{dT} = t \rho_e \int_{E_{\nu_{min}}} \frac{d\sigma_{\rm DP}}{dT}\ \frac{d\Phi (\nuebar)}{dE_\nu} dE_\nu\,, 
\end{equation}
%**********************************FIGURES**************************************************
where $\rho_e$ is electron number density per kg of the target mass, t is data taking period and $d\Phi/dE_\nu$ corresponds to neutrino spectrum. For various $M_{A^\prime}$ values, a minimum $\chi^2$ fit is applied to find the 90\% CL limits for the coupling constant $\gbl$ by defining it in the following form 
$$
\chi^2 = \sum \limits_{i=1} \frac{[R_{\rm Exp}(i) - (R_{\rm SM}(i) + R_{\rm DP}(i))]}{\Delta_{\rm Stat}(i)}\,,
$$
where $R_{\rm SM}(i)$ and $R_{\rm DP}(i)$ are the expected event rate on the $i^{th}$ data bin due to SM and DP contributions, respectively, and $\Delta_{\rm Stat}(i)$ is the corresponding uncertainty in the measurement.

\begin{figure}[h]
	$\begin{array}{cc}
	\includegraphics[width=16cm]{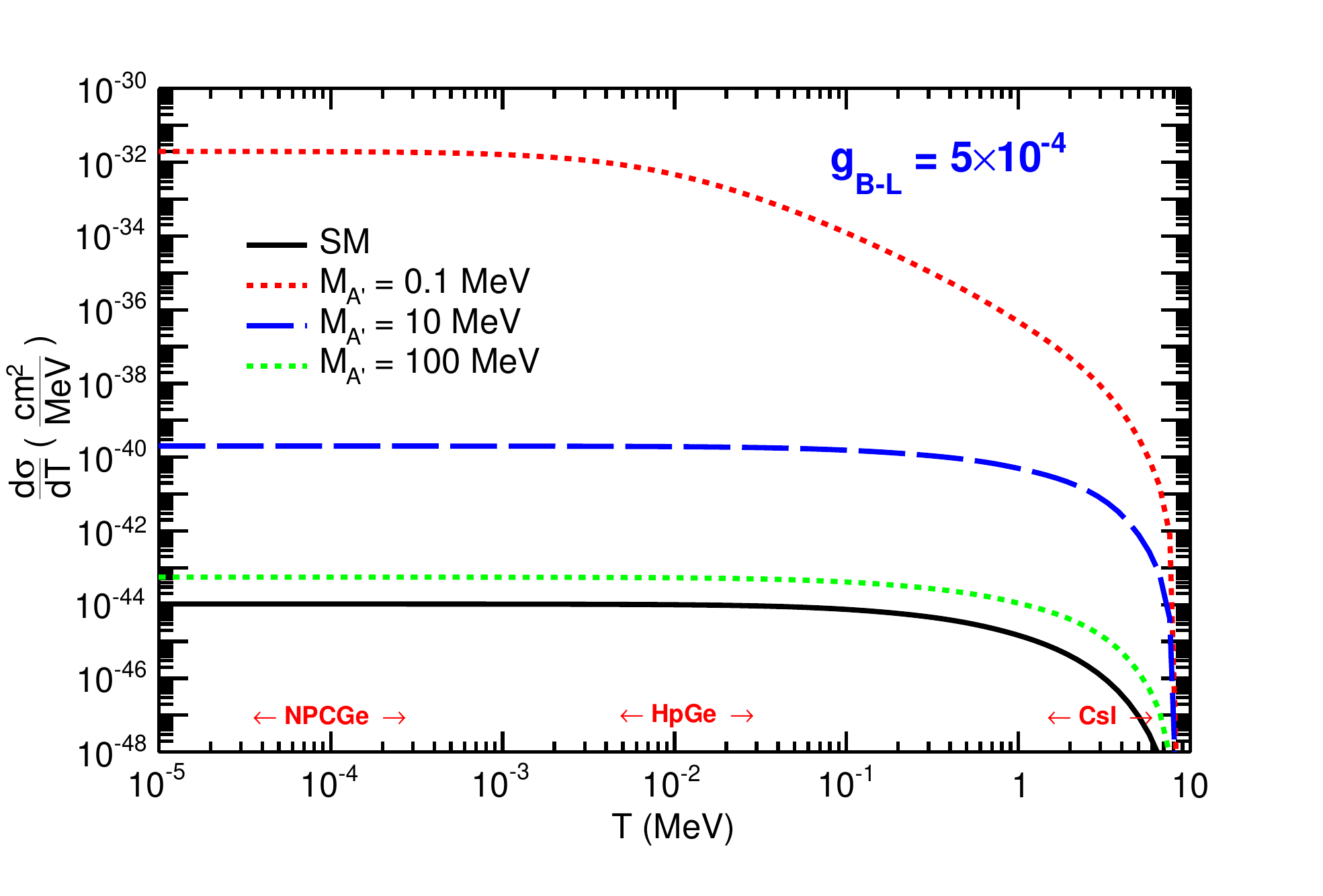}
	\end{array}$
	\vspace*{-0.7cm}
	\caption{\label{cross_section_vs_recoil} Cross-section vs recoil for various $M_{A^\prime}$ by normalizing neutrino flux to 1.}
\end{figure}
Let us analyse how differential cross section at a fixed value of $\gbl$ changes as a function of $T$ for various $M_{A^\prime}$ values in Fig.~\ref{cross_section_vs_recoil}. Rhe chosen value for $\gbl$ is just representative. For larger $M_{A^\prime}$ values like 0.1 MeV, 10 MeV or 100 MeV, only larger $T$ tail of the differential cross section has some $T$ dependency but it becomes flat when $T$ gets smaller than $M_{A^\prime}$. This is expected since as $T$ gets much smaller than $M_{A^\prime}$, the factor $\left(M_{A^\prime}^2 + 2mT\right)^{-2}\rightarrow 1/M_{A^\prime}^4$ and in addition to this the other factor in the cross section expression is dominated by $E_\nu$ in the small $T$ region. Overall, a flat profile is obtained. The point where the curves start being flat moves to smaller recoil energy T values as smaller and smaller $M_{A^\prime}$ values are taken. High sensitivity of the differential cross section to $M_\aprime$, which in turn gives better bounds of $\gbl$ is another motivation for searching very light dark photon through $\nu-e$ scattering experiments.

\subsection{Roles of Interference}
\label{int}

\begin{figure}[b]
	$\begin{array}{cc}
	\hspace{-0.7cm}		\includegraphics[scale=0.46,keepaspectratio]{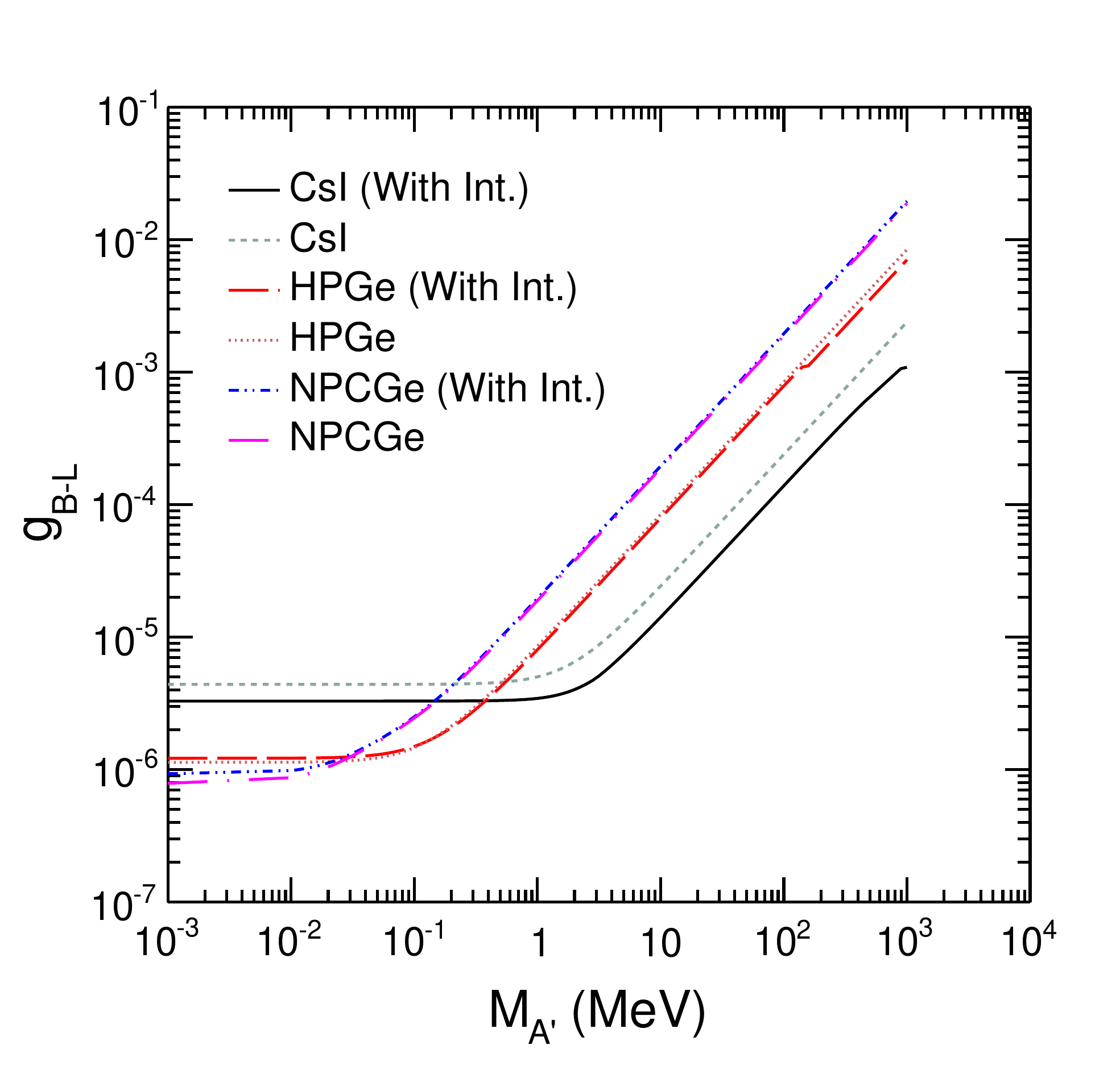} &
	\hspace{-0.8cm}		\includegraphics[scale=0.46,keepaspectratio]{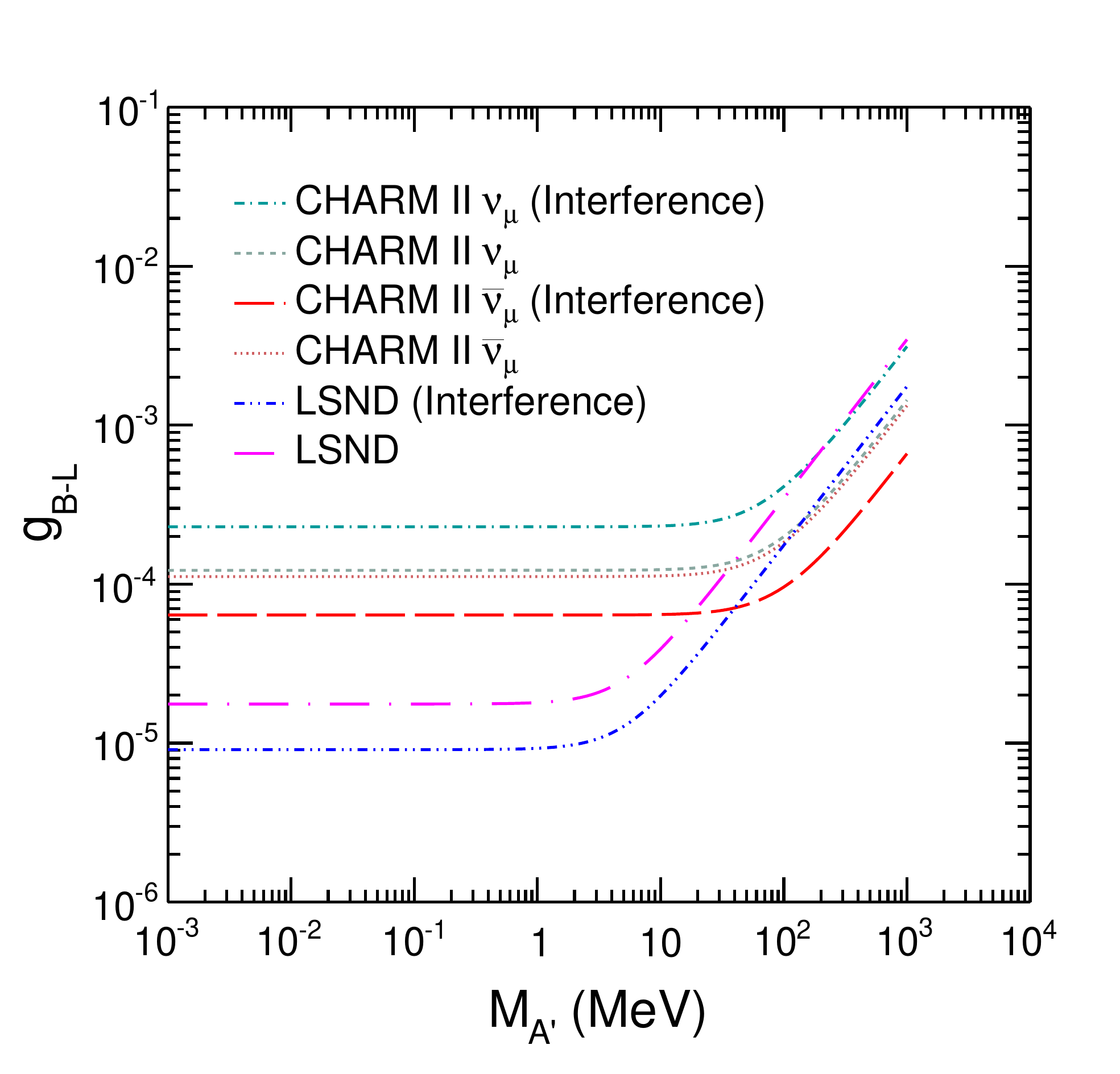} \\
	(a)    & (b) 
	\end{array}$
	\caption{\label{texono_with_without_int} The $90\%$ CL exclusion limits in the $g_{\rm B-L} - M_{A^\prime}$ plane for various TEXONO experiments in the panel (a) and for the LSND and CHARM II  experiments in the panel (b). The results with and without the interference contributions are shown for highlighting its significance.}
\end{figure}

A theme of  this study is to explore the roles of interference effects between contributions from new physics and SM.  The interference between the SM and new physics contribution due to vector boson exchange in neutrino-electron scattering processes are illustrated in Figs.~\ref{texono_with_without_int}a and \ref{texono_with_without_int}b, showing the exclusion limits for $\gbl$ versus $M_{A^\prime}$ in TEXONO experiments ($\bar{\nu}_e$ at O(1 MeV)) together with LSND ($\nu_\mu$ at O(10 MeV)) and CHARM II ($\nu_\mu$ and  $\bar{\nu}_\mu$ O(10 GeV)) experiments. 

 It can be seen from Fig.~\ref{texono_with_without_int} that although for low recoil energies (T $\sim$ keV) interference term does not affect the bound on coupling constant $\gbl$, there is an enhancement in general due to interference for higher recoil energy values (T $\sim$ MeV). Contributions of interference terms are sizable when the effects due to new physics  are small relative to the SM contributions. This is the case applicable to experiments where the SM cross-sections are measured, such as LSND, TEXONO-Csl, CHARM II and BOREXINO, where the interference effects on the parameters $\gbl$ and $M_{A'}$ are depicted in Fig.~\ref{texono_with_without_int}. Otherwise, when the ranges of new physics effects are large compared to SM, the interference term can in general be neglected.

The interference effects between SM and new physics due to dark photons can be both constructive or destructive. As seen from Fig. \ref{texono_with_without_int}, the interference is destructive only in the  $\numu$ electron scattering case of the CHARM II experiment. In all other cases, the interference is constructive so that more stringent bounds can be derived. The behavior of CHARM II result can be seen from Eqn.~(\ref{sigmaint}) where the differential cross sections take the following forms 
\begin{eqnarray}
\frac{d\sigma_{\rm INT}(\nu_\alpha e^-)}{dT}\, &\propto&\  T(T-2E_\nu)\,, \\ 
\frac{d\sigma_{\rm INT}(\bar{\nu}_\alpha e^-)}{dT}\,   &\propto&\ - T(T-2E_\nu)\,,
\end{eqnarray}
with $\s2tw\simeq 1/4$. 
In general, $T/2< E_{\nu_{min}}$ such that the interference terms are always positive (constructive) or negative (destructive) for  $\bar{\nu}_\alpha$ ($\nu_{\alpha}$), respectively.  
A similar analysis can be done for the $\nu_e$ and $\bar{\nu}_e$ scatterings, where the interference is constructive.

	\begin{figure}
\hspace{-0.8cm}		\includegraphics[scale=0.6]{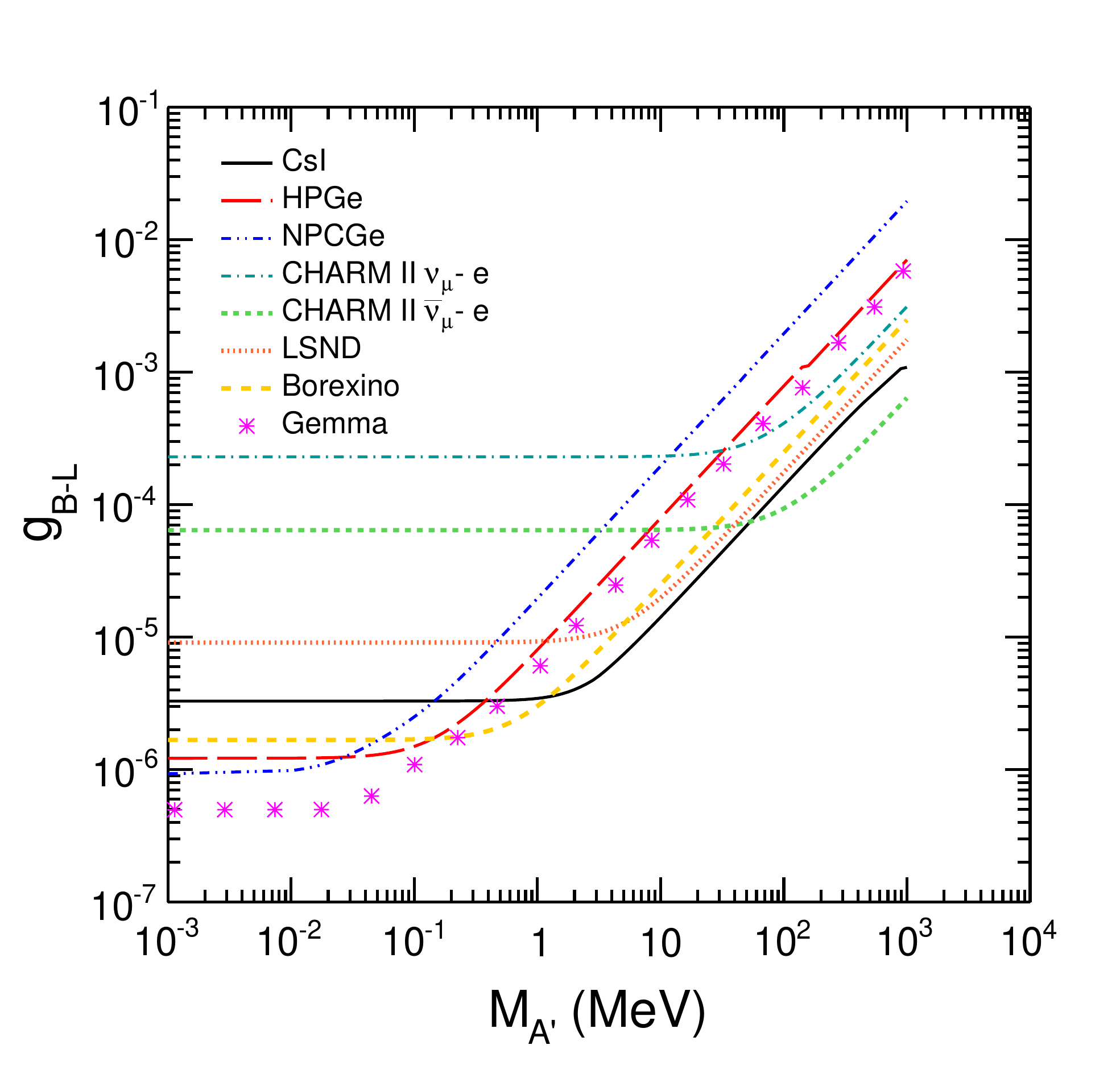}
		\vspace*{-0.6cm}
		\caption{\label{Int_result} The $90\%$ C.L. exclusion limits of the gauge coupling constant $g_{\rm B-L}$ of the $U(1)_{\rm B-L}$ group as a function of the dark photon mass $M_{A^\prime}$ by including interference effects. The regions above the curves are excluded.}
	\end{figure}

\subsection{Results}
\label{results}	
With interference effects properly accounted for, the exclusion limits in the $M_{A^\prime}-g_{\rm B-L}$ plane including all relevant neutrino-electron scattering  experiments are shown in Fig. \ref{Int_result}. The BOREXINO results \cite{Harnik:2012ni} with interference are included, provided better bounds by about $30 \%$. It was verified that switching off the interference term would reproduce those of Ref.~\cite{Harnik:2012ni}. Best limits for different parts of exclusion regions come from different reactor neutrino experiments; by GEMMA and TEXONO-CsI for $M_{A'}< 0.1$ MeV and $0.1<M_{A'}<100$ MeV, respectively, whereas by accelerator neutrinos data from CHARM II ($\bar{\nu}_\mu$) for $M_{A'}>100$ MeV.

The behavior of the exclusion curves of Fig.~\ref{Int_result} can be understood through the dark photon cross section expression of Eqn.~(\ref{pure_DP_cs}), with a dependence of $(M_{A'}^2 + 2mT)^{-2}$ . Accordingly, studies of dark photons favor experiments with low energy neutrinos like those from reactors. At  $M_{A'}\ll T$, cross section is insensitive to $M_{A'}$, implying that (i) neutrino-electron scattering experiments would not be able to resolve dark photons with mass less than keV, which is the lower reach of current sensitivities on $T$; (ii) accelerator experiments with $E_\nu$ and $T$ at the GeV range would not provide good sensitivities, except at $M_{A'}$ also larger than GeV . 
\begin{figure}[h]
	\hspace*{-1.0cm}\includegraphics[width=19cm]{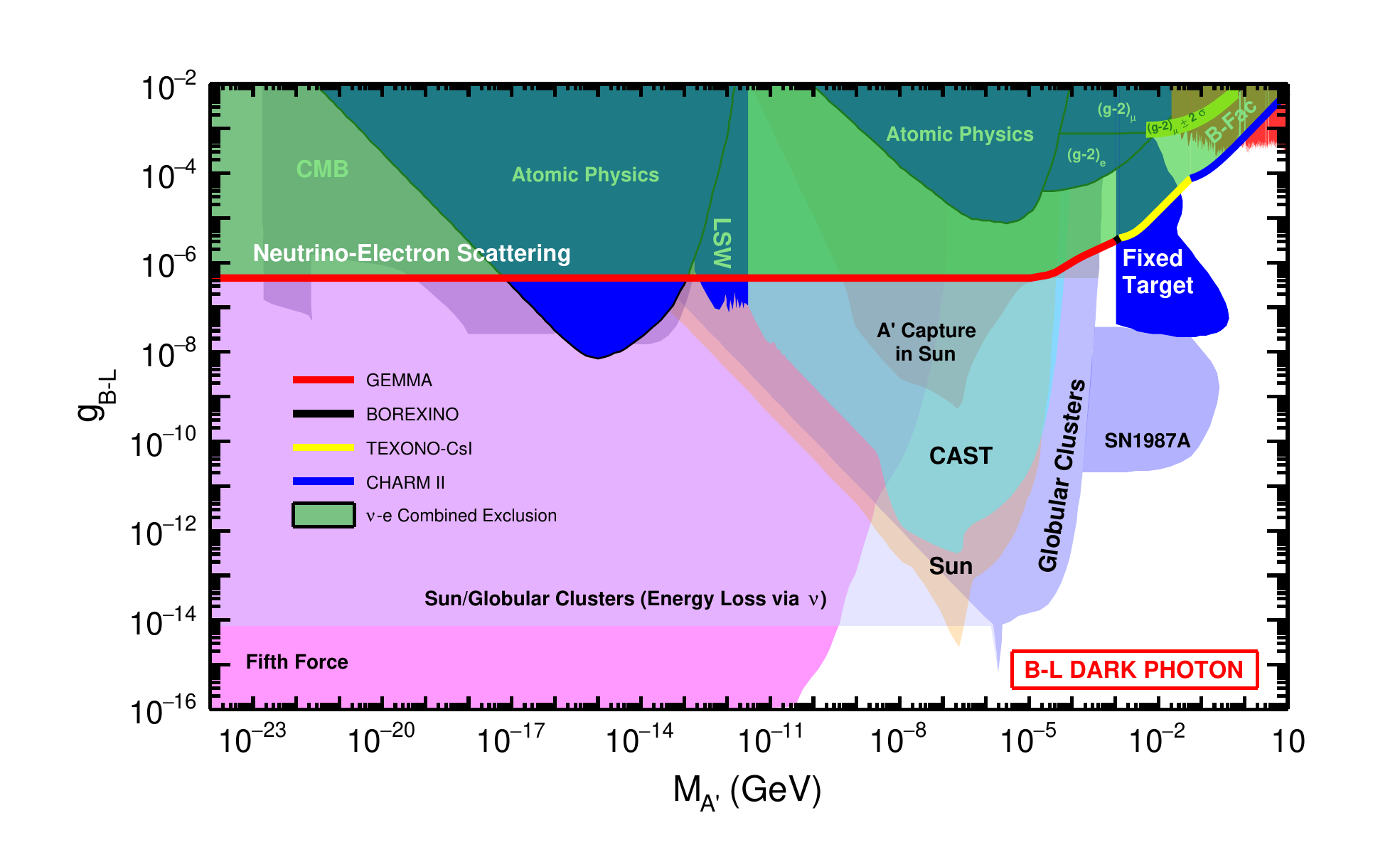}
	\vspace*{-1.4cm}
	\caption{\label{paramspace-Aprime} The current global exclusion plot of the bounds on the gauge coupling of the dark photon from different cosmological and astrophysical sources as well as laboratory experiments, adapted from Ref.~\cite{Harnik:2012ni} with combined limits from neutrino-electron scattering (this work) overlaid. The $90\%$ C.L. bounds are defined by the GEMMA, BOREXINO, TEXONO-CsI, CHARM II ($\bar{\nu}_\mu$) experiments, from low to high $M_{A^\prime}$. The two white regions in large $M_{A^\prime}$ above the exclusion line are the new parameter space probed and excluded by $\nu-e$ scattering experiments. The principles of the different categories of experiments are summarized in Table \ref{tab::others}.}
\end{figure}

Exclusion regions from the $\nu-e$ scattering experiments are displayed with other laboratory and cosmological bounds in Fig.~\ref{paramspace-Aprime}, which corresponds to an update of Fig. 8a in Ref.~\cite{Harnik:2012ni}  with recent data. The robust bounds are shown with dark color shading, while those involving assumptions, considered less robust, are with lighter transparent shading. The basic methods in these different categories of experiments are summarized in Table III.

Few comments on the recent data in Fig.~\ref{paramspace-Aprime} are in order. The data of the other laboratory and cosmological bounds which are plotted on the kinetic mixing $\epsilon$ and $M_\aprime$ plane can be used to constrain $\gbl$. The conversion is $\epsilon \to \frac{(B-L)(f)}{Q_f}\gbl$ for experiments not involving  decay modes of the dark photon. Otherwise the conversion would have the additional factor $\left(\frac{BR(\aprime\xrightarrow{\small B-L} f\bar{f})}{BR(\aprime\xrightarrow{\epsilon} f\bar{f})}\right)^{\!\!1/2}$ in the right hand side for experiments involving $\aprime$ decaying to fermions. 

The excluded regions labeled as ``Sun" and ``Globular Clusters" are originally presented in Refs.~\cite{Redondo:2008aa,Raffelt:1988rx}. Recently, it has been shown in Ref.~\cite{An:2013yfc}\footnote{We thank R.Harnik for bringing this point to our attention.} that emission of the forgotten longitudinal modes of the dark photon change the stellar constraints drastically in especially small $M_{A'}$ region. Consequently, the region excluded by the so-called light-shinning-through-wall (LSW) experiments falls under the tails of the excluded regions from the stellar bounds \cite{An:2013yfc}. 

Exclusions from the recent BaBar data \cite{Lees:2014xha}, marked ``B-Fac" in Fig.~\ref{paramspace-Aprime}, covers a wider $M_\aprime$ region. The $2 \sigma$ allowed band from the muon $g-2$ experiment \cite{Bennett:2006fi} is also shown. Part of the allowed band ($M_\aprime\gtrsim 0.02$ GeV) is rejected by ``B-Fac" data (see also the proposal \cite{Kahn:2014sra} to probe the lower $M_\aprime$  regions). Results from neutrino-electron scattering experiments also probe and exclude that region by an order of magnitude. There is an alternative scenario discussed in \cite{Lee:2014tba} that a $U(1)_L$ gauge boson may survive all constraints so that the  $(g-2)_\mu$ allowed band may remain compatible with other data.

It can be seen that Figs.~\ref{Int_result} and \ref{paramspace-Aprime} that the bounds on $\gbl$ from $\nu-e$ scattering experiments are insensitive to $M_\aprime$ at $M_\aprime < 10$ keV. The constraints are not expected to change drastically since it is experimentally challenging to measure even lower recoil energies. At  $10^{-3}\, {\rm eV} \lesssim M_\aprime \lesssim 1\, {\rm MeV} $,  the $\nu-e$ data significantly improve the current bounds if only the robust limits are adopted. Two new regions are probed at $4 \times 10^{-4}\, {\rm GeV} \lesssim M_\aprime \lesssim 10^{-3}$ GeV and $ 10^{-2}\,{\rm GeV} \lesssim M_\aprime \lesssim 1$ GeV over currently published results. However, the sharp cutoff at $2m_e$ for ``Fixed Target" experiments is based on the channel $\aprime \to e^+e^-$. The gap at $M_\aprime \simeq 10^{-3}$ GeV is expected to be probed when invisible channels like $\aprime \to 2\nu$ in the case of $\rm B-L$ dark photon would be taken into account.

\begin{table*}[t]
	\caption{\label{tab::others} The list of different sources used to bound on the gauge coupling of the dark photon. The details of each are summarized very briefly together with references for the details. }
	\begin{ruledtabular} 
		\begin{tabular}{p{2.05cm} p{10cm} p{2.1cm}}
			%				\hline
			\textbf{Experiments}		& \centering{\textbf{Comments}}   &\textbf{References} \\
			\hline
			g-2			&  $\aprime$ contribution to magnetic moment of $e$ and $\mu$.  & \cite{Pospelov:2008zw,Bennett:2006fi}\\
			\hline
			Fixed Target			& %\begin{minipage} {0.6\textwidth} %\begin{itemize}\itemsep-0.4cm
					$A^\prime$ production in beam dump experiments.
					$\aprime \rightarrow e^- e^+$ in $M_\aprime > 2m_e$.
				%\end{itemize}
			%\end{minipage}    
			& \cite{Bjorken:2009mm,Batell:2009di,Essig:2010gu,Izaguirre:2014dua,Diamond:2013oda,Abrahamyan:2011gv}          \\
			\hline
			B-Factories			& %\begin{minipage}{0.6\textwidth} \begin{itemize}\itemsep-0.4cm 
					 $\Upsilon \rightarrow \gamma \aprime$ and $\aprime \rightarrow \gamma\ l^+ l^-.$ Sensitive to range $0.02\, {\rm GeV} < M_\aprime < 10.2$ GeV.
				%\end{itemize}
			%\end{minipage}
			&\cite{Essig:2009nc,Carone:1994aa,Graesser:2011vj,Lees:2014xha}                 \\
			\hline
			Fifth Force			& %\begin{minipage}{0.6\textwidth} \begin{itemize}\itemsep-0.4cm 
					Precision measurements of gravitational, Casimir and Van der Waals forces.
					Sensitive to $M_ \aprime \lesssim 100~\rm eV$.
			%	\end{itemize}
			%\end{minipage}
			&\cite{Jaeckel:2010ni,Adelberger:2006dh}  \\
			\hline
			Atomic Physics			&  Corrections to Coulomb Force.  &  \cite{Jaeckel:2010ni,Bartlett:1988yy} \\
			\hline
			%\textbf{Astrophysical Bounds}\\
			
			Supernova			& Analysis of energy loss of Supernova.   & \cite{Dent:2012mx,Kazanas:2014mca}   \\
			\hline
			Sun			&  Luminosity analysis in the conversion of plasmons in the sun.   & \cite{Redondo:2008aa,An:2013yfc, Raffelt:1988rx}  \\
			\hline
			LSW			&  Transition of  $laser \rightarrow \aprime \rightarrow \gamma$.    & \cite{Jaeckel:2010ni,Ahlers:2007qf}    \\
			\hline
			CMB			&   Study of black body spectrum of Cosmic Microwave Background.   &  \cite{Jaeckel:2010ni,Mirizzi:2009iz}    \\
			\hline
			CAST			&  Comparison of flux of dark and usual photon.        & \cite{Redondo:2008aa,Arik:2008mq}                      \\
			\hline
			Globular Clusters  & Energy loss due to dark photons in Globular Clusters.  & \cite{Redondo:2008aa,An:2013yfc, Raffelt:1988rx,Jaeckel:2010ni} \\
			%		\hline
		\end{tabular}
	\end{ruledtabular} 
\end{table*}

%\clearpage
\section{Conclusions}
\label{conc}

A very light dark photon from hidden sector through a vector portal could couple to some SM particles which might give a signal via neutrino electron scattering experiments if, especially, the dark photon is the gauge field of a $U(1)$ group gauged with $B-L$ symmetry. Indeed this will allow a direct coupling with neutrinos, which modifies the electroweak contribution with a presumed negligible interference. The new interactions due to existence of $\aprime$ boson whose couplings do not contain derivatives lead to differential cross section being proportional to $~1/T^2$ which makes low energy neutrino experiments sensitive to dark photon search in the low mass region. Hence low energy neutrino experiments like TEXONO which aims to measure neutrino nucleus coherent scattering as well as neutrino magnetic moment has advantage to search new gauged boson, located much below the electroweak scale. For the higher mass region for $\aprime$ boson, neutrino experiments with higher incident energy have better sensitivity. 

We have done a study to search for the signal of dark photon originating from a $U(1)_{\rm B-L}$ group in the available data sets of the TEXONO as well as GEMMA, BOREXINO, LSND, and CHARM II. With no signal, our analysis is converted to a bound on the gauge coupling $\gbl$ as a function of $M_\aprime$. One of the crucial part of our study is to look at the interference between the dark photon diagram and the ordinary photon  in detail. Our results show that the interference effects are significant for experiments with a smaller deviation from the SM prediction. Other than the CHARM II $\nu_\mu$ electron scattering case, all the others have constructive interference which makes the bounds more stringent. The BOREXINO case where the interference effects are sizable are updated. Our results consolidate and expand the excluded regions in particular probing new parameter space $4 \times 10^{-4}\, {\rm GeV} \lesssim M_\aprime \lesssim 10^{-3}$ GeV and ${\rm 10^{-2}\, GeV \lesssim}\ M_{A^\prime}\ {\rm \lesssim 1\, GeV}$. The recent BaBar data gives better bound for $M_\aprime$ bigger than 1 GeV. The $2\sigma$ favorable band of $(g-2)_\mu$ which is partly excluded by BaBar for rather heavier $M_\aprime$, the remaining low mass region is also excluded by our results.
 
The experimental bounds would not be improved significantly by the future neutrino experiments since  the pure new physics differential cross section is proportional to the fourth power of coupling constant $\gbl^4$.

\begin{acknowledgements}
I.T. and S.B. thank  METU-BAP grant number 08-11-2013-028. I.T. also thanks TUBA-GEBIP for its partial support. S.B. acknowledges TUBITAK-2211 program for its partial support. T.M.A. thanks the Visiting Professor program at King Saud University for its partial support. We thank R.Harnik for sending us the pdf version of Fig.~8a in Ref.~\cite{Harnik:2012ni}. We are grateful G.Krnjaic for useful correspondences and B.Echenard for providing us the recent BaBar data.  I.T. and S.B are thankful to Altug Ozpineci for bringing \LaTeX~TikZ package to their attention. 
\end{acknowledgements}

%\nocite{*}
%\bibliographystyle{plain}
%\bibliography{dp}
\end{document}